\documentclass[sigconf, nonacm]{acmart}

\definecolor{todocolor}{RGB}{200,0,0}      

\newcommand{\sparagraph}[1]{\vspace{1mm}\noindent\textbf{#1}}

\usepackage{enumitem}
\setlist[itemize]{leftmargin=15pt}

\usepackage{etoolbox}
\usepackage{xcolor}

\usepackage{graphicx}
\usepackage{multirow}

\newcommand\vldbworkshop{Accelerating Analytics and Data Management Systems Using Modern Processor and Storage Architectures}

\newcommand\vldbyear{2026}

\newcommand\vldbauthors{\authors}
\newcommand\vldbtitle{\shorttitle}

\newcommand\vldbavailabilityurl{}

\newcommand\vldbpagestyle{plain}

\begin{document}

\title{High-Throughput LLM-Based Compression and What It Takes}
\subtitle{A Systems Study of Inference Engines and Entropy Coding Optimizations}

\author{David Adams}
\orcid{}
\affiliation{%
  \institution{University of Melbourne}
  \city{Melbourne}
  \country{Australia}
}
\email{david.adams1@student.unimelb.edu.au}

\author{Luisa Neubauer}
\orcid{}
\affiliation{%
  \institution{University of Technology Nuremberg}
  \city{Nuremberg}
  \country{Germany}
}
\email{luisa.neubauer@utn.de}

\author{Ping-Lin Kuo}
\orcid{}
\affiliation{%
  \institution{University of Technology Nuremberg} 
  \city{Nuremberg}
  \country{Germany}
}
\email{ping-lin.kuo@utn.de}

\author{Mihail Stoian}
\orcid{}
\affiliation{%
  \institution{University of Technology Nuremberg}
  \city{Nuremberg}
  \country{Germany}
}
\email{mihail.stoian@utn.de}

\author{Thomas R\"{u}ckstie\ss}
\orcid{}
\affiliation{%
  \institution{Relaxed Constraints}
  \city{Melbourne}
  \country{Australia}
}
\email{thomas@relcon.ai}

\author{Josif Grabocka}
\orcid{}
\affiliation{%
  \institution{University of Technology Nuremberg}
  \city{Nuremberg}
  \country{Germany}
}
\email{josif.grabocka@utn.de}

\author{Renata Borovica-Gajic}
\orcid{}
\affiliation{%
  \institution{University of Melbourne}
  \city{Melbourne}
  \country{Australia}
}
\email{renata.borovica@unimelb.edu.au}

\author{Andreas Kipf}
\orcid{}
\affiliation{%
  \institution{University of Technology Nuremberg}
  \city{Nuremberg}
  \country{Germany}
}
\email{andreas.kipf@utn.de}

\begin{abstract}
Large language models are attractive text compressors because next-token probabilities can be converted into code lengths. In practice, however, the main bottleneck is often not modeling quality but systems cost: a compressor must obtain logits quickly enough to compete with strong classical baselines.

This paper examines systems optimizations for LLM-based compression, including multistream pipelined arithmetic coding and target-interval capture, and evaluates their trade-offs across inference engines, configurations and datasets. We evaluate LLM-based text compression on three different inference engines: Transformers, vLLM, and TensorRT-LLM. Notably, we obtain a speedup of 18.84x compared to prior work's Transformers implementation and a maximum recorded inference speed of 181 kilobytes per second at a compression ratio of 7.17, when running on an NVIDIA A100.
\end{abstract}

\maketitle

\pagestyle{\vldbpagestyle}
\begingroup\small\noindent\raggedright\textbf{VLDB Workshop Reference Format:}\\
\vldbauthors. \vldbtitle. VLDB \vldbyear\ Workshop: \vldbworkshop.\\ 
\endgroup
\begingroup
\renewcommand\thefootnote{}\footnote{\noindent
This work is licensed under the Creative Commons BY-NC-ND 4.0 International License. Visit \url{https://creativecommons.org/licenses/by-nc-nd/4.0/} to view a copy of this license. For any use beyond those covered by this license, obtain permission by emailing \href{mailto:info@vldb.org}{info@vldb.org}. Copyright is held by the owner/author(s). Publication rights licensed to the VLDB Endowment. \\
\raggedright Proceedings of the VLDB Endowment. 
ISSN 2150-8097. \\
}\addtocounter{footnote}{-1}\endgroup

\ifdefempty{\vldbavailabilityurl}{}{
\vspace{.3cm}
\begingroup\small\noindent\raggedright\textbf{VLDB Workshop Artifact Availability:}\\
The source code, data, and/or other artifacts have been made available at \url{\vldbavailabilityurl}.
\endgroup
}


\section{Introduction}\label{sec:introduction}

Real-world analytical database systems store and process substantial amounts of textual data. In an analysis of the Amazon Redshift fleet,
\texttt{varchar} columns account for 52.1\% of stored columns, making variable-length strings the most prominent datatype in the observed workloads \citep{vanrenenWhyTPCNot2024}. Similarly, an analysis of production Snowflake workloads reports that text columns account for 34\% of aggregation inputs, substantially more than other non-numeric types \citep{szlangWorkloadInsightsSnowflake2025}. These findings suggest that efficient storage and processing of textual data are relevant concerns for modern analytical database systems.

Compression is particularly attractive for textual data that is retained but accessed or modified infrequently. The Redshift study reports that 86.6\% of fleet tables received no inserts or deletes during the monitored period, and identifies compression as a potential optimization for such data \citep{vanrenenWhyTPCNot2024}. It also finds that half of the queries reading external data sources target compressed files, nearly all of which use gzip, demonstrating that compression already forms part of production analytical data pipelines. For cold text data, stronger compression could reduce storage and data movement costs, provided that it does not impose prohibitive costs when data is ingested or subsequently decoded for analysis.

Large language models (LLMs) offer a new approach to lossless compression of
textual data.
In LLM-based compression, a pretrained model supplies next-token probability distributions, which an entropy coder converts into a compact bitstream. 
While these distributions may enable substantial storage reductions, obtaining them requires neural inference during both compression and decompression.
Thus, to be practically useful in accelerating data management systems, LLM-based compression must be evaluated against core Database Management System (DBMS) metrics. 
Specifically, we must measure the trade-offs between storage footprint reduction, ingestion throughput, and decompression speed during query execution.

\sparagraph{Motivation.} This motivates our central question: under which combinations of inference engine, hardware platform, and entropy-coding design might LLM-based compression become economically viable as a storage codec for textual data? To investigate this question, we evaluate the systems-level economics of deploying LLM-based compressors across inference engines and their configurations, and entropy coder optimizations, examining their trade-offs and viability alongside standard DB compressors FSST \cite{fsst} and \texttt{zstd} on these crucial dimensions.


\sparagraph{Related Work.} Neural lossless compression treats compression as next-token prediction followed by entropy coding. Early work by \citet{schmidhuberSequentialNeuralText1996} showed that neural sequence models could be combined with Huffman \cite{Huffman1952AMF} or arithmetic coding \cite{rissanenArithmeticCoding1979} for lossless text compression, establishing the basic connection between neural prediction quality and conventional text compression algorithms. Later systems used deeper neural models to improve compression ratios. DeepZip \cite{goyalDeepZipLosslessData2019} combines gated RNNs, including LSTMs and bidirectional GRUs, with arithmetic coding and reports compression ratio improvements over gzip on text and genomic data. DecMac \cite{liuDecMacDeepContext2019} similarly uses an LSTM-based context model with cycle connections, reporting further reductions over gzip and 7zip. These systems show that neural predictors can improve on compression ratio, but also make compression substantially more expensive because model training and token-by-token inference dominate runtime.

DZip \cite{goyalDZipImprovedGeneralpurpose2021} reduces some of this cost with a hybrid adaptive training strategy, avoiding separate external training data while remaining general-purpose, but still reflects the central trade-off in neural compression: better probability estimates require more computation. TRACE \cite{maoTRACEFastTransformerbased2022} shifts this trade-off toward faster inference by replacing recurrent models with a lightweight transformer architecture. Its single-layer transformer, byte grouping, and shared feed-forward components target the sequential bottlenecks of earlier neural compressors. 

Recent work replaces these task-trained predictors with pre-trained large language models. LLMZip \cite{llmzip} demonstrates this idea by using a pre-trained LLaMA-7B \cite{touvronLLaMAOpenEfficient2023} model, reporting strong compression ratios on English text. \citet{lossless_data_compression_llm} extend this argument more broadly, showing that large pre-trained models can serve as powerful compressors across different data modalities. 

However, removing training does not remove the systems cost. LLM-based compression still requires repeated next-token probability queries, often in a highly sequential loop. FineZip \cite{finezip} and Nacrith \cite{tacconelliNacrithNeuralLossless2026} therefore focus on making LLM-based compression more practical through online memorization, lightweight auxiliary predictors, higher-precision coding, backend optimizations and selective use of the LLM. \citet{Kipf2026WaitingTD} explore the trade-offs between compression ratios and latency and make the limitations explicit from a systems perspective: the compression gains of LLMs must be weighed against slow inference and decompression costs.

A parallel line of work studies the problem of serving LLMs efficiently. vLLM \cite{kwonEfficientMemoryManagement2023} improves throughput by managing the key-value cache using PagedAttention, reducing memory waste and enabling larger effective batch sizes. SGLang \cite{zhengSGLangEfficientExecution2024} introduces runtime optimizations such as RadixAttention and efficient structured decoding for programs that make many language model calls. Although these systems were not designed primarily for compression, they are highly relevant because LLM-based arithmetic coding is dominated by repeated probability queries. 

\sparagraph{Contribution.} This work sits at the intersection of neural compression and inference systems. Prior LLM-based compression work argues that stronger predictive models can improve code length, but often treats the cost of obtaining logits as a secondary concern~\cite{llmzip,finezip,lossless_data_compression_llm,compression_llm_generated_data}.

We also relate to work on high-throughput LLM inference. Engines such as vLLM, Transformers, and TensorRT-LLM focus on different inference optimizations, and those design choices can directly affect whether compression is viable
on server GPUs, CPUs, or even edge devices. 

Finally, we compare against strong classical baselines such as LZ4 and xz. These compressors remain the practical reference point because they are fast, cheap, and broadly deployed. Our goal is therefore not only to ask whether LLM-based compression can improve compression ratio, but under which system configurations it can do so at acceptable speed and cost~\cite{boncz20,adaptivecompression,mosaic}.

In particular, our optimized implementation on these three inference engines achieves speedups of 18.84x, 1.85x, and 2.16x, respectively, over prior work's baseline implementation~\cite{Kipf2026WaitingTD}, when run on a NVIDIA A100, making a step towards high-throughput LLM-based compression.
\section{Problem Setting}\label{sec:problem-setting}

We implement lossless compression as an append-only next-token prediction workload. Given a tokenized sequence, the compressor repeatedly queries a language model on the current prefix and converts the predicted distribution for the next token into compressed output. This workload makes logit access, prefix reuse, and serving overhead more important than general inference.

\subsection{Global-Mask Compression}

Let $x_{1:n}$ be the token sequence obtained from the input under the tokenizer
of a fixed language model $M$. The implementation first constructs a global
allowed-token set
\[
  V' =
  \begin{cases}
    \{x_i : 1 \le i \le n\} \subset V, & \text{if vocabulary reduction is enabled},\\
    V, & \text{otherwise},
  \end{cases}
\]
where $V$ is the full tokenizer vocabulary. The set $V'$ is serialized as a
bitmap and stored with the compressed payload. All subsequent probability
vectors are restricted to this single global vocabulary, rather than using a
different candidate set per chunk.

The token stream is split into $B$ contiguous rows of lengths $m_1,\dots,m_B$,
where $B$ is the configured batch size, and the first token of each row is
stored verbatim as the seed needed for decoding. At step $t$, the compressor
forms the context
\[
  c_t = x_{\max(1,\,t-R+1):t},
\]
where $R$ is the configured context-length limit. The predictor queries the
model for next-token logits
\[
  z_t = M(c_t) \in \mathbb{R}^{|V|}.
\]
For probability-based encodings the logits are restricted to $V'$ and
normalized:
\[
  p_{t+1,\,j} =
  \frac{\exp z_{t,j}}{\displaystyle\sum_{k \in V'}\exp z_{t,k}},
  \qquad j \in V'.
\]
Arithmetic coding encodes the true next token $x_{t+1}$ under the probability
mass function $p_{t+1}$, consuming $-\log_2 p_{t+1,\,x_{t+1}}$ bits, so the
model's cross-entropy directly determines the ideal compressed size before
coding overhead and the bitmap.

The implementation also supports rank-based encodings. In that case the
predictor returns logits rather than probabilities, and the compressor records
the zero-indexed rank
\[
  r_{t+1} = \bigl|\{j \in V' \mid z_{t,j} > z_{t,\,x_{t+1}}\}\bigr|.
\]
The rank sequence is then encoded with either a fixed-width bitpacked
representation or a Huffman code. These encodings measure how often the model
places the correct token near the top of the distribution, while avoiding a
full probability-to-arithmetic-coder conversion at each step.

Decompression mirrors the same autoregressive process. Given the bitmap,
initial row tokens, and compressed payload, the decoder reconstructs $V'$,
queries the same model on the reconstructed prefixes, and decodes each next
token from the arithmetic stream. The decoded token is appended to the context
for the following query. Compression is lossless only when the model,
tokenizer, vocabulary mask, context policy, and arithmetic-coding parameters
are identical between compression and decompression.
\section{Implementations}\label{sec:implementations}

Due to the nature of modern LLM inference engines and the compression workload, we found it necessary to implement custom logit capture paths for both vLLM and TensorRT-LLM. These paths are not intended for general use, but they allow us to evaluate the trade-offs of different inference engines without confounding factors from separate compressor implementations. The same compressor loop can be used with Hugging Face (HF)'s Transformers, vLLM, or TensorRT-LLM backends, and the same predictor interface can be used for custom logit capture or engine-backed scores.

\subsection{Custom and Engine-Backed Logits}

The baseline path uses HF's Transformers directly. It pads
variable-length prompts when needed, forwards the prompt batch through the
model, takes the final-position logits, optionally restricts them to the global
token mask, and applies a softmax only for probability-based encodings. With KV
caching enabled, the first call builds a cache from the full prompt and
subsequent calls feed only the newly appended token until the context is
truncated and the cache must be rebuilt. This path is simple and gives explicit
control over logits, but each compression step is still a Python-level loop
around model execution, score conversion, and entropy coding.

Because vLLM is optimized for text generation, its batch scheduler does not
return next-token scores in the row-stable order required for compression. We
therefore use a POSIX shared-memory interface: a logits processor runs inside
the vLLM worker after the dense logits are computed, copies the selected scores
into a pre-allocated capture buffer, and marks each row-step entry with a
readiness flag. Each request carries a row identifier so the host can recover
the original batch order regardless of vLLM's internal scheduling. When
vocabulary reduction is enabled, only the reduced-vocabulary columns are
copied, keeping transfer costs proportional to $|V'|$ rather than the full
padded vocabulary. Implementation details and earlier design iterations are
described in Appendix~\ref{app:engine-backends}.

The TensorRT-LLM path uses the runtime's offline APIs, which expose generation
logits directly. The predictor builds or loads a cached engine configured for
the compression batch size and context length, issues a one-token generation
request per prompt batch, and receives a dense $[\mathrm{batch},\mathrm{vocab}]$
tensor of final-position scores. KV-cache management is handled internally by
the runtime, so no host-side cache maintenance is required. The same global
token mask and deferred softmax used by the other backends are applied before
scores are passed to the entropy coder.

\subsection{Deployment Variants}

We evaluate three practical deployment styles.

\sparagraph{Transformers backend.} This backend loads the Hugging Face model in
the compressor process. It supports LoRA adapters, seq2seq and Mamba variants
where configured, and explicit cache-free or cached operation. It is the most
portable implementation and serves as the correctness reference for the other
engines.

\sparagraph{vLLM backend.} This backend uses vLLM's offline
\texttt{LLM.generate} API with prefix caching enabled when requested. It
supports tensor parallelism, configurable GPU memory utilisation, and a fixed
maximum model length matching the compression context length. Generation is used
as the mechanism that causes vLLM to run the model and invoke the logits
processor; the generated token is not the primary output.

\sparagraph{TensorRT-LLM backend.} TensorRT-LLM is integrated through the same
predictor interface. It can build or reuse a cached engine directory and return
the same reduced probability/logit matrices as the other paths. The
implementation currently targets decoder-only models and relies on
TensorRT-LLM's internal runtime cache management. This keeps the evaluation
axes clean: model, hardware, inference engine, and encoding can vary without
changing the compressor loop.

\subsection{Optimization}

Using unmodified engines with arithmetic coding is not economical
\cite{Kipf2026WaitingTD}. We frame three optimizations not just as compressor tweaks, but as generic acceleration patterns for LLM-as-a-service within databases. These patterns target different parts of the end-to-end critical path: request scheduling inside the inference engine, proba\-bility-to-interval conversion for arithmetic coding, and score-tensor transfer back to the host.

\sparagraph{Teacher-forced windows.} A naive compressor issues one
next-token request per compression step, paying the fixed overhead of request
creation, scheduling, and shared-memory synchronisation at every token. Instead,
the compressor sends the current prefix together with the next $w$ ground-truth
tokens from the input stream. The vLLM logits processor records the score vector
at each internal decode step and forces the sampler to emit the known target
token, so one engine call collects scores for an entire window. This amortises
per-request overhead across $w$ compression steps and allows the KV cache to be
reused across successive positions within the window. See
Appendix~\ref{app:optimizations} for compatibility caveats and measured
overheads.

\sparagraph{Multistream pipelined compiled arithmetic coding.} Because each
batch row is compressed independently, the implementation maintains one range
coder per row and converts probability vectors to target intervals in tensor
form, avoiding per-row Python loops. A Numba-compiled~\cite{Lam2015NumbaAL} backend
executes the integer range-coder update loop over compact arrays of interval
triples. On the Transformers backend, interval conversion and encoding for one
batch are overlapped with the next model forward pass via a background thread,
reducing exposed CPU-side coding latency. Full quantization and pipelining
details are given in Appendix~\ref{app:optimizations}.

\sparagraph{Target-interval capture.} The arithmetic coder requires only the
quantized interval assigned to the true next token, not the full probability
vector. In target-interval mode the inference backend computes this interval in
place and returns a fixed-size record $(\ell, h, T, p_{\mathrm{target}})$ per
row and step. For the vLLM backend this is particularly effective: the logits
processor performs the softmax and interval derivation inside the worker and
writes only the fixed-size record to shared memory, reducing transfer cost from
$O(|V'|)$ floats to $O(1)$ per row-step.
\section{Experiments}
\label{sec:experiments}

All experiments use the \texttt{Qwen/Qwen2.5-0.5B} model. With the exception of the hardware section, all experiments were run on the same hardware
configuration: two AMD EPYC Milan 7313 16-core 3\,GHz processors, 512\,GB DDR4-3200 ECC memory, and a NVIDIA A100 40\,GB SXM4 4-GPU module. The default operating point is a batch size of 256, a context length of 256, and 100 retained tokens from the dataset (as the initial context). We utilize datasets from prior work~\cite{Kipf2026WaitingTD}: (i) the \texttt{text8} dataset (100\,MiB), an English text extracted from Wikipedia; (ii) Python code, namely the first 100\,MiB from PyTorrent~\cite{pytorrent}; and (iii) the Yelp dataset, containing 150,346 business records from the Yelp Open Dataset.\footnote{Data obtained from \texttt{https://www.yelp.com/dataset}.} Reported throughput is in MB/s (uncompressed bytes per wall-clock second) and compression ratio as original size divided by compressed size. 

\subsection{Experiment A: Engine Comparison}
\label{sec:exp-a}

Table~\ref{tab:engine-comparison} compares the three inference engines at a fixed batch size and encoding configuration. The transformer model is also evaluated without its KV cache to isolate the cost of the incremental forward pass.

\begin{table}[h]
\caption{Inference engine comparison.}
\label{tab:engine-comparison}
\begin{tabular}{llccc}
\toprule
\textbf{Engine} & \textbf{Variant} &  \textbf{(MB/s)} & \textbf{ratio} & \textbf{speedup} \\
\midrule
\multirow{2}{*}{Transformers}
  & + KV cache            & 0.0344 & 7.0492 & $1.00\times$ \\
  & No KV cache           & 0.0030 & 7.0502 & $0.09\times$ \\
\midrule
vLLM
  & Default               & 0.0130 & 7.0304 & $0.37\times$ \\
\midrule
\multirow{2}{*}{TensorRT-LLM}
  & Batch 256             & 0.0081 & 7.0499 & $0.023\times$ \\
  & Batch 512 (max)       & 0.0065 & 7.9370 & $0.019\times$ \\
\end{tabular}
\end{table}

The most striking result in Table \ref{tab:engine-comparison} is that neither of the two production-grade inference engines outperforms the plain Hugging Face's Transformers. Specifically, vLLM reaches only $0.37\times$ the Transformers' throughput and TensorRT-LLM fares worse still at $0.023\times$. This is not a flaw in the engine but a mismatch between their design objectives and the compression workload. They are designed for high-concurrency serving: they optimize request scheduling, memory pooling, and latency fairness across independent queries. LLM-based AC, by contrast, is a stateful, sequential workload in which every token depends on the one before it and all rows in a batch must be processed in the correct order to feed the entropy coder. As the per-request overhead of this workload is negligible in a serving context the scheduling, shared-memory synchronization and result reassembly becomes the dominant cost incurred by these engines. 

\begin{figure}
    \centering
    \includegraphics[width=1\linewidth]{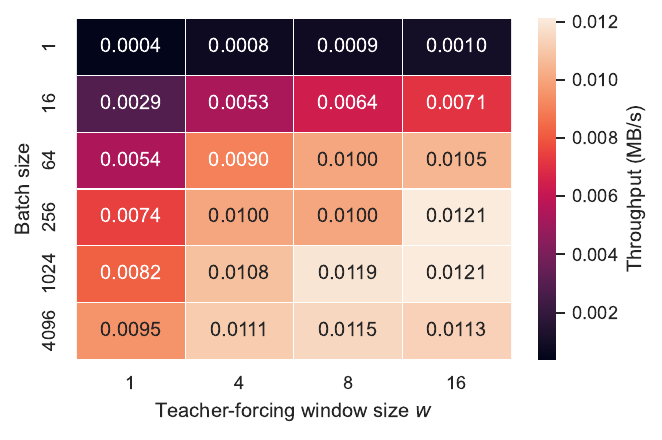}
    \caption{vLLM throughput (MB/s) as a function of both batch size and teacher-forcing window size $w$ (varying O1, using O2). Setting $w=1$ corresponds to no teacher forcing.}
    \label{fig:vllm-window-sweep}
\end{figure}

\subsection{Experiment B: Optimization Ablation}
\label{sec:exp-b}

We evaluate every applicable combination of the three
compression-specific optimizations:
\begin{itemize}
  \item \textbf{O1.} Teacher-forced vLLM windows (vLLM only; window size $w=8$).
  \item \textbf{O2.} Multistream pipelined arithmetic coding
        (Multistream AC, Numba packed backend and pipeline encoding on HF's Transformers).
  \item \textbf{O3.} Target-interval capture (Target interval AC).
\end{itemize}
Combinations that are architecturally inapplicable (O1 on transformer or TensorRT-LLM) are marked N/A.  Compression ratio is expected to remain constant across all rows within an engine because O2 and O3 are lossless algorithmic equivalents of the baseline. 

\begin{table*}[t]
\centering
\caption{Optimization ablation at batch size 256, context length 128, \texttt{text8} dataset.  Relative speedup is normalized to each engine's own baseline.  O1 applies only to vLLM.}
\label{tab:optimization-ablation}
\small
\begin{tabular}{llcccc}
\toprule
\textbf{Engine} & \textbf{Configuration}
  & \textbf{(MB/s)}
  & \textbf{ratio}
  & \textbf{Rel.\ speedup}
  & \textbf{AC time (\%)} \\
\midrule
\multirow{4}{*}{Transformer}
  & Baseline (AC)         & 0.0018 & 7.0482 & $1.00\times$ & 97.1 \\
  & + O2                  & 0.0018 & 7.0482 & $1.00\times$ & 97.1 \\
  & + O3                  & 0.0347 & 7.0492 & $18.96\times$ & 1.0 \\
  & + O2 + O3             & 0.0345 & 7.0492 & $18.84\times$ & 1.0 \\
\midrule
\multirow{8}{*}{vLLM}
  & Baseline (AC)         & 0.0031 & 7.03 & $1.00\times$ & 89.9 \\
  & + O1                  & 0.0116 & 7.05 & $3.43\times$ & 9.2 \\
  & + O2                  & 0.0116 & 7.05 & $3.43\times$ & 9.2 \\
  & + O3                  & 0.0065 & 7.03 & $1.94\times$ & 3.0 \\
  & + O1 + O2             & 0.0118 & 7.05 & $3.48\times$ & 9.2 \\
  & + O1 + O3             & 0.0063 & 7.05 & $1.87\times$ & 2.3 \\
  & + O2 + O3             & 0.0066 & 7.03 & $1.94\times$ & 3.1 \\
  & + O1 + O2 + O3        & 0.0062 & 7.05 & $1.85\times$ & 2.3 \\
\midrule
\multirow{4}{*}{TensorRT-LLM}
  & Baseline (AC)         & 0.0020 & 7.0488 & $1.00\times$ & 58.2 \\
  & + O2                  & 0.0020 & 7.0488 & $1.00\times$ & 58.2 \\
  & + O3                  & 0.0044 & 7.0499 & $2.18\times$ & 0.7 \\
  & + O2 + O3             & 0.0043 & 7.0499 & $2.16\times$ & 0.7 \\
\bottomrule
\end{tabular}
\end{table*}

Table \ref{tab:optimization-ablation} exposes a fundamental bottleneck-shift pattern that differs sharply across engines, and which has direct implications on how each engine should be configured in a database storage stack. 

\paragraph{Transformers.}

At baseline, the Transformers engine spent 97.1\% of wall time inside the arithmetic coder. In this case the model forward pass is essentially free by comparison. O2 alone achieves no improvement because without O3 there is still a full probability vector to process per step and the pipeline has nothing useful to overlap. O3 alone eliminates the bottleneck entirely: by computing only the quantized interval for the true next token inside the inference step, the AC time drops from 97.1\% to 1.0\% and throughput 
increases by $18.96\times$. This is the single most impactful optimization across all engines. Adding O2 on top of O3 yields minimal gains for such small datasets  as there is not enough work to adequately distribute across threads. 

\paragraph{vLLM}

The vLLM baseline is slightly more balanced than the Transformers' 89.9\% AC time, but achieves only a slightly higher absolute baseline of 0.0031\,MB/s, reflecting the per-request overhead that is always present regardless of coding cost. O1 (teacher forced windows) and O2 produce identical speedups of $3.43\times$ and their AC profiles are also similar. This suggests that both reduce the same cost of issuing one engine call per token but via different mechanisms and they are not additive. O3 alone
gives a more modest $1.94\times$ (0.0065\,MB/s), because unlike the Transformers case, inference overhead is not negligible in vLLM, so eliminating the coding step does not produce a proportionally large gain.

Crucially, combining all three optimisations (O1+O2+O3) slightly \emph{regresses} to 0.0062\,MB/s ($1.85\times$). This counter-intuitive result reflects a scheduling interaction: O1 restructures requests into windows that the vLLM scheduler processes differently, and when O3 is also active the reduced shared-memory transfer per step interacts with the window batching in a way that slightly increases idle time.

\paragraph{TensorRT-LLM}

TensorRT-LLM shows a markedly different AC time profile at baseline:
58.2\%, indicating that the engine itself already imposes substantial non-coding overhead compared to the plain Transformers. O2 yields no improvement, as with Transformers. O3 delivers a $2.18\times$ speedup (0.0044\,MB/s), but this is far smaller than the $18.96\times$ seen on Transformers because a larger fraction of time is already spent in engine overhead that O3 cannot address. This suggests that TensorRT-LLM's internal runtime costs of engine initialization, offline API scheduling, and KV-cache management present a floor that no coding-side optimization can break through. Unless future versions of TensorRT-LLM expose lower-overhead logit access paths for stateful sequential workloads, it is unlikely to be competitive with the Transformers backend for this use case.

\subsection{Experiment C: Batch-Size Sweep}
\label{sec:exp-c}

We report in Figure~\ref{fig:vllm-window-sweep} the throughput at every batch size for vLLM, making the bottleneck shift visible, and additionally sweeping the teacher-forcing window sizes $w \in \{1, 4, 8, 16\}$ to quantify the O1 $\times$ batch interaction. 
It shows that the benefit of teacher-forcing is most pronounced at small-to-medium batch sizes, where the per-request scheduling overhead dominates. At larger batch sizes the inference step itself becomes more GPU-saturating, and the marginal value of amortizing request overhead across a window of $w$ tokens diminishes. This mirrors a well-known pattern in database query execution: batching reduces fixed per-tuple overheads but offers diminishing returns once the variable per-tuple cost dominates. From an operator's perspective, window size $w=8$ represents a practical sweet spot: it delivers most of the available gain without the additional memory pressure of larger windows.

\begin{figure}
    \centering
    \includegraphics[width=1\linewidth]{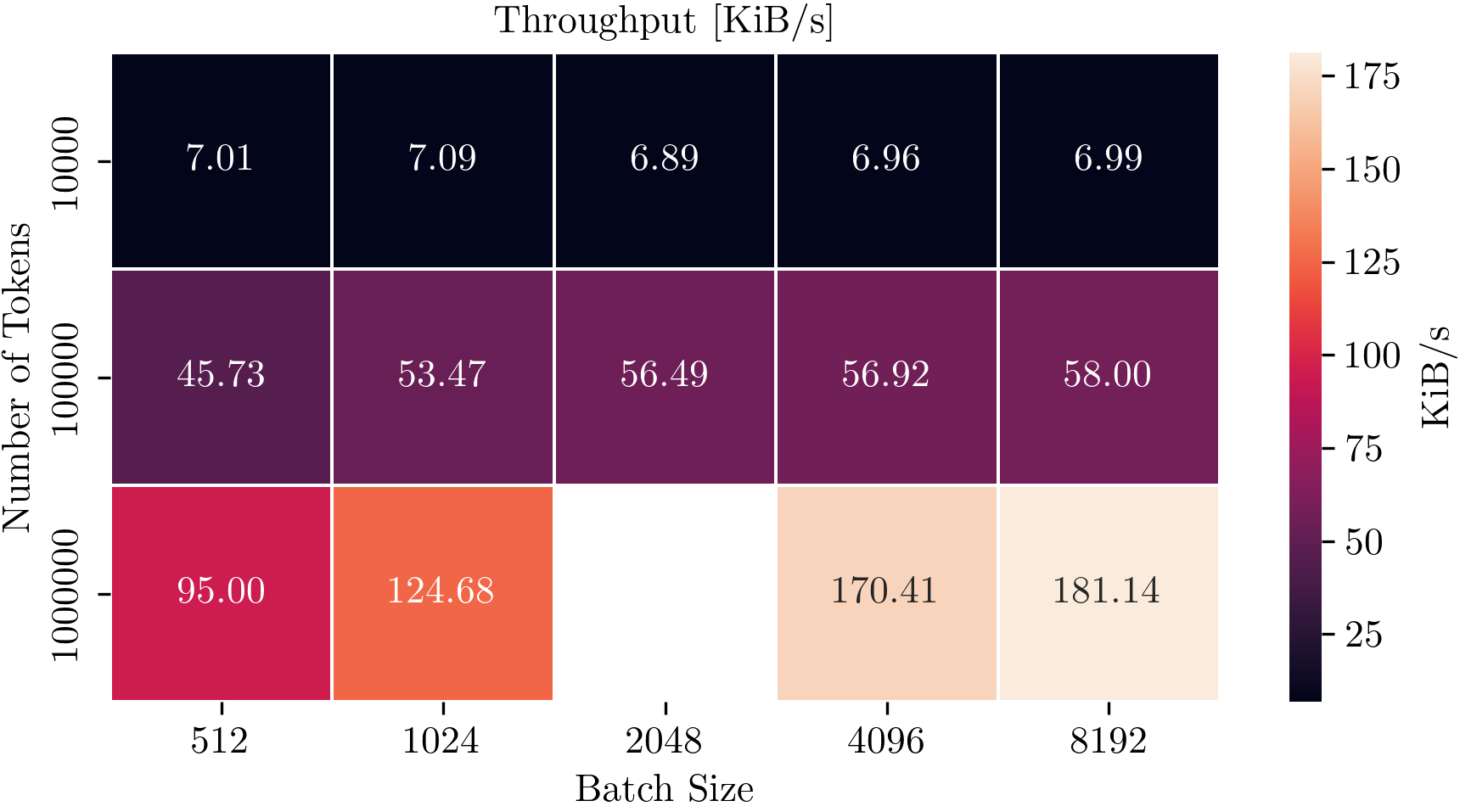}
    \caption{Transformers throughput (Kb/s) as a function of both batch size and number of tokens processed. Missing values are repeating out of memory errors.}
    \label{fig:throughput-sweep}
\end{figure}

Figure \ref{fig:throughput-sweep} shows that processing more tokens improves overall hardware utilization and throughput.
 
\subsection{Experiment D: Context-Length Sensitivity}
\label{sec:exp-d}

Longer contexts increase the per-step inference cost and alter KV-cache reuse patterns.  Table~\ref{tab:context-sweep} reports throughput and compression ratio for Transformers at the fixed batch size across context lengths $R \in \{32, 64, 128, 256, 512\}$ with retain length set to $R/2$.

\begin{table}[H]
\centering
\caption{Context-length sensitivity on Transformers with the fully optimized encoding path (+ O2 + O3), batch size 256, \texttt{text8} dataset. Retain length is set to half the context length in each case. The default operating point (context 256, retain 100) is shown in bold.}
\label{tab:context-sweep}
\begin{tabular}{cccc}
\toprule
\textbf{Context length} & \textbf{Retain tokens} & \textbf{(MB/s)} & \textbf{ratio} \\
\midrule
32  & 16  & 0.0324 & 6.2288 \\
64  & 32  & 0.0338 & 6.6592 \\
128 & 64  & 0.0339 & 6.9239 \\
\textbf{256} & \textbf{100} & \textbf{0.0337} & \textbf{7.0600} \\
512 & 256 & 0.0349 & 7.1007 \\
\bottomrule
\end{tabular}
\end{table}

Table~\ref{tab:context-sweep} presents a practically important
trade-off for storage system designers. Throughput is nearly flat across the range 64-512 tokens (0.0338-0.0349\,MB/s), varying by less than 3\%. Compression ratio, however, improves materially with context length: from 6.23 at context 32 to 7.10 at context 512. The improvement is concave: the largest gains occur between context lengths 32 and 128, with diminishing returns thereafter consistent with the model's ability to exploit local lexical and syntactic patterns within the first hundred tokens, beyond which marginal predictive value from additional history decreases.

This decoupling of throughput from context length is a favorable property from a systems perspective. A storage administrator can increase context length to improve the compression ratio of cold data reducing storage footprint and data movement costs at essentially no throughput penalty, provided GPU memory is sufficient to hold the enlarged KV cache. At context 512 the KV cache is twice as large as at context 256 but delivers only a 0.7\% throughput improvement (7.10 vs.\ 7.06 ratio), making 256 the more memory-efficient default. 

\subsection{Experiment E: Encoding Quality}
\label{sec:exp-e}

Table~\ref{tab:encoding-quality} reports compression quality for each encoding scheme on the text8 dataset with the Transformers backend at the default operating point. Arithmetic-coding variants (AC, Multistream AC, Target interval AC) and PMATIC are lossless-optimal up to quantisation; rank-based schemes (bitpacked, Huffman) trade compression ratio for coding simplicity.  Classical baselines are included for reference.

\begin{table}
\centering
\caption{Compression quality on \texttt{text8}, Transformers engine, batch size 256, context length 256.  Bits/token is the empirical cross-entropy; classical baselines do not use an LLM so bits/token is not applicable.}
\label{tab:encoding-quality}
\small
\begin{tabular}{lccc}
\toprule
\textbf{Encoding} & \textbf{(MB/s)} & \textbf{ratio} & \textbf{Bits/token} \\
\midrule
AC (Baseline)         & 0.0359 & 6.1962 & 0.0000 \\
Multistream AC (O2)   & 0.0345 & 7.0492 & 5.1642 \\
Target interval AC (O3) & 0.0349 & 7.1007 & 5.1201 \\
PMATIC                & timeout & -- & -- \\
Bitpacked             & 0.0363 & 2.8735 & N/A \\
Huffman               & 0.0359 & 6.1962 & N/A \\
\midrule
\multicolumn{4}{l}{\textit{Classical baselines (CPU, no LLM)}} \\
\midrule
LZ4                   & 666.66 & 1.83 & N/A \\
xz -9e                & 1.34 & 4.30 & N/A \\
zstd                  & 586.70 & 3.06 & N/A \\
FSST                  & 294.11 & 2.08 & N/A \\
\bottomrule
\end{tabular}
\end{table}

\begin{table}
\caption{Compression quality on different datasets, fully optimized Transformers engine (+ O2 + O3), batch size 256, context length 256.}
\label{tab:encoding-datasets}
\begin{tabular}{lcc}
\toprule
\textbf{Dataset} & \textbf{{(MB/s)}} & \textbf{ratio} \\
\midrule
\texttt{text8}  & 0.005436 & 7.739  \\
\texttt{code}   & 0.004132 & 11.203  \\
\texttt{Yelp}   & 0.002490 & 7.028  \\
\bottomrule
\end{tabular}
\end{table}

Table \ref{tab:encoding-quality} places the LLM-based encodings in context against alternative coding schemes and classical compressors. Among the arithmetic-coding variants, Target interval AC (O3) achieves the best ratio (7.1007) at 0.0349\,MB/s, marginally improving on Multistream AC (7.0492 at 0.0345\,MB/s). The difference
in bits/token 5.1201 vs.\ 5.1642 is small but consistent, reflecting that target-interval capture loses slightly less to quantisation by operating directly on the coder interval rather than on a rounded probability vector.

The rank-based schemes tell a cautionary tale. Bitpacked encoding retains the throughput of arithmetic coding but collapses the ocmpression ratio to 2.87, worse than zstd (3.06) at a fraction of the cost. Huffman encoding matches AC in tatio but offers no throughput advantage. PMATIC's timeout is notable: it represents an approach that attempts higher coding precision at the cost of per-token latency that is not practical. 

\subsection{Discussion of DB-Centric Compressors}
\label{sec:db-centric-compressors}

When placing these DB-centric classical baselines (e.g., zstd, FSST) on the throughput-ratio cost Pareto frontier, we observe that they heavily prioritize ingestion throughput and decompression speed. While LLM-based compression achieves unprecedented compression ratios by accurately modeling the data distribution, its throughput remains orders of magnitude slower, fundamentally bottlenecked by the model's forward pass. In scenarios querying cold storage where storage cost dominates, LLM compression may be viable. However, for active query paths and high-throughput ingestion, lightweight compressors like FSST and zstd currently provide a far more economical balance.

\subsection{Experiment F: Encoding Backend Ablation}
\label{sec:exp-f}

This experiment holds the inference engine (Transformers with KV cache) and batch size fixed while varying only the arithmetic-coding backend, thread count, and pipelining flag.  The goal is to isolate coding overhead from inference cost and to guide backend selection across hardware targets.

The four Numba-capable encodings (AC, Multistream AC, Target interval AC) are each tested with the \texttt{python}, \texttt{numba}, \texttt{numba\_threaded}, and \texttt{numba\_packed} backends. For threaded backends, thread counts of 1, 4, and auto (0, which selects the number of CPU cores) are included.  For \texttt{Multistream AC} and \texttt{Target interval AC}, pipelining (\texttt{--pipeline\_encoding}) is tested both on and off.  Rank-based encodings (\texttt{bitpacked}, \texttt{huffman}) and \texttt{PMATIC} are included as single reference rows since they do not expose a backend choice.

Table~\ref{tab:backend-ablation-multistream} covers Multistream AC;
and Table~\ref{tab:backend-ablation-target} covers Target interval AC

\begin{table}[t]
\centering
\caption{Backend $\times$ pipeline ablation for O2 (Multistream AC),
  Transformers engine, batch size 256, \texttt{text8} dataset.}
\label{tab:backend-ablation-multistream}
\small
\begin{tabular}{lllccc}
\toprule
\textbf{Backend} & \textbf{Threads} & \textbf{Pipeline} & \textbf{(MB/s)} & \textbf{AC (\%)} & \textbf{speedup} \\
\midrule
Raw python      & --   & off & 0.0008 & 36 &   $1\times$ \\
numpy          & --   & off & 0.0914 & 13.3 & $57.01\times$ \\
numba\_threaded & auto & off & 0.0920 & 9.3 & $57.42\times$ \\
numba\_threaded & 4    & off & 0.0935 & 9.7 & $58.36\times$ \\
numba\_threaded & 1    & off & 0.0940 & 9.4 & $58.67\times$ \\
numba\_packed   & --   & off & 0.0949 & 9.2 & $59.22\times$ \\
numba\_threaded & auto & on  & 0.4847 & 48.9 & $302.48\times$ \\
numba\_threaded & 4    & on  & 0.4860 & 49.1 & $303.29\times$ \\
numpy           & --   & on  & 0.4867 & 71.0 & $303.71\times$ \\
numba\_packed   & --   & on  & 0.4879 & 47.1 & $304.45\times$ \\
numba\_threaded & 1    & on  & 0.4879 & 49.2 & $304.49\times$ \\
\bottomrule
\end{tabular}
\end{table}

\begin{table}[t]
\centering
\caption{Backend $\times$ pipeline ablation for O3 (Target interval AC), Transformers engine, batch size 256, \texttt{text8} dataset.}
\label{tab:backend-ablation-target}
\small
\begin{tabular}{lllccc}
\toprule
\textbf{Backend} & \textbf{Threads} & \textbf{Pipeline} & \textbf{(MB/s)} & \textbf{AC (\%)} & \textbf{speedup} \\
\midrule
Raw python      & --   & off & 0.0008 & 36 &   $1\times$ \\
numpy           & --   & off & 0.0908 & 13.4 & $56.63\times$ \\
numba\_threaded & 4    & off & 0.0936 & 9.6 & $58.42\times$ \\
numba\_threaded & auto & off & 0.0938 & 9.4 & $58.52\times$ \\
numba\_threaded & 1    & off & 0.0940 & 9.4 & $58.65\times$ \\
numba\_packed   & --   & off & 0.0951 & 9.1 & $59.35\times$ \\
numba\_threaded & 1    & on  & 0.4808 & 48.4 & $300.05\times$ \\
numpy           & --   & on  & 0.4813 & 71.2 & $300.34\times$ \\
numba\_threaded & 4    & on  & 0.4836 & 48.6 & $301.77\times$ \\
numba\_threaded & auto & on  & 0.4851 & 48.8 & $302.69\times$ \\
numba\_packed   & --   & on  & 0.4853 & 46.5 & $302.85\times$ \\
\bottomrule
\end{tabular}
\end{table}

Tables~\ref{tab:backend-ablation-multistream}
and~\ref{tab:backend-ablation-target} isolate the arithmetic coding
layer from the inference engine and reveal two distinct operating
regimes separated by the pipelining flag.

\paragraph{Without pipelining.}
The jump from a simple python implementation of AC with cumulative multiplications to a vectorized numpy implementation is the largest single step: $57\times$ for Multistream AC. This reflects the cost of per-token Python interpreter overhad in the inner coding loop: at batch size 256 there are 256 independent range coders to update per step which is trivially parallelisable in numpy. Numba just in time compilation offers further gains of 2-4\% over numpy. At this scale threading provides no benefit: numba\_threaded at 1, 4, and auto threads all land within 1\% of each other (0.0920--0.0940\,MB/s for Multistream AC). This indicates that the bottleneck without pipelining is not thread-level parallelism but the memory bandwidth of updating 256 compact interval arrays, a pattern that saturates a single core.

Enabling the pipeline flag causes a further ${\sim}5\times$ jump across all backends, converging to 0.48--0.49\,MB/s regardless of whether numpy, numba, or numba\_packed is used. The tight clustering spread across all pipelined backbends show that pipelineing the active coding step is fully overlapped with model inference and the remaining AC time fraction (41-71\%) show that the inference is to short to fully cover the remining work. 

The practical implication here is that the choice of using numba matters only when pipelining is unavailable e.g a CPU-only deployment. In such a regime numba\_packed is the best single choice. When pipelining is available,
numpy suffices and avoids the just in time compilation overhead at startup, which may be relevant for short-lived compression jobs.

\paragraph{With pipelining.}

Enabling the pipeline flag causes a further ${\sim}5\times$ jump across all backends, converging to 0.48--0.49\,MB/s regardless of whether numpy, numba, or numba\_packed is used. The tight clustering shows that with pipelining active the coding step is fully overlapped with model inference and the remaining AC time fraction (47--71\%) reflects
coding work that cannot be hidden because the inference step is too
short to cover it. The Numpy backend shows the highest AC time fraction (71.0\%) at comparable throughput, suggesting slightly less efficient overlap than Numba backends, but
the absolute difference in throughput is negligible.

The practical implication is the choice of Numba variant matters only when pipelining is
unavailable (e.g., on a CPU-only deployment): in this regime,
numba\_packed is the best single choice. When pipelining is available, Numpy suffices and avoids the just in time compilation overhead at startup, which may be relevant for short-lived compression jobs.



\subsection{Reproducibility Notes}

To assess run-to-run variability, we repeated a subset of throughput measurements (Transformers engine, no optimizations, varying batch sizes and context lengths) five times per configuration and found low variation, with a mean coefficient of variation at 0.76\%. Due to limited compute resources, we therefore report single-run measurements for the remaining experiments.
\section{Discussion: Throughput Bottleneck Analysis}\label{sec:bottlenecks-outline}

The end-to-end compressor has three potential bottlenecks: inference
latency, arithmetic-coding pre-processing, and score-extraction overhead. 
The dominant bottleneck shifts as batch size and engine optimizations change.

\sparagraph{Low-batch regime.}
At small batch sizes, throughput is determined by model inference latency.
The compressor stalls after each token step waiting for the next score vector,
so faster engines such as vLLM and TensorRT-LLM directly raise throughput by
shortening that wait. GPU utilization is low in this regime because fixed
launch costs dominate and the arithmetic-coding side is idle most of the time.

\sparagraph{Arithmetic-coding pre-processing at larger batches.}
As batch size grows, the arithmetic-coding side becomes the bottleneck.
The baseline AC path requires probability normalization, conversion to
integer frequencies, and cumulative-prefix construction for every active
row at every compression step $O(|V'|)$ work repeated across the full
batch. This overwhelms the savings from larger GPU batches and caps useful
parallelism.

\sparagraph{Effect of multistreamed arithmetic coding and pipelining.}
The multistream backend reduces this cost in two ways. First, it eliminates
full cumulative-vector construction per token event: target intervals are
staged compactly and fed directly to row-local range coders. Second,
the Numba-compiled backends move the hot renormalization loop out of Python
and, in the packed variant, write byte-oriented streams directly without
materializing intermediate bit lists. When pipelining is enabled on the
transformer backend, interval staging and encoding are overlapped with the
next model forward pass. Together these changes remove enough exposed
coding latency that larger GPU batch sizes become useful again, shifting
the bottleneck back toward inference.

\sparagraph{Score-extraction overhead.}
Once arithmetic-coding cost is reduced, the next bottleneck for the vLLM
and transformer paths is score extraction. Both still require a dense
logit or reduced-vocabulary probability matrix to reach the host at
every step unless target-interval capture is active. For vLLM, this
transfer is particularly expensive because logits must traverse shared
memory out of the worker process, and that cost can dominate short
next-token requests. Target-interval capture mitigates this by reducing
the exported payload to a constant-size record per row-step, but
dense-logit capture remains costly when full distributions are needed.

\sparagraph{Transformer backend after optimization.}
Despite being less specialized than TensorRT-LLM and less
throughput-oriented than vLLM, the transformer backend is the most
flexible path for this workload: it exposes logits directly, supports
host-managed control flow, and can overlap arithmetic coding with the
next forward pass without shared-memory capture hooks. As a result, once
the AC path is optimized, the transformer backend can match or exceed the
other engines on end-to-end compression throughput even when those engines
are faster for ordinary text serving. This reflects a broader point: the
compression workload has different systems requirements from generation,
and inference engines designed with direct interval or reduced-vocabulary
capture in mind could reduce all three bottlenecks simultaneously.
\section{Conclusion}\label{sec:conclusion}

This paper evaluated the systems-level economics of LLM-based lossless text compression across inference engines, entropy-coding optimisations, and hardware configurations. While LLMs consistently achieve strong compression ratios over $7\times$ on natural language and $11\times$ on code, the key question is whether they can do so economically compared to classical compressors.

Our results show that the main bottleneck is not model inference but arithmetic coding. High-throughput inference engines such as vLLM and TensorRT-LLM perform poorly in this setting because their scheduling strategies are mismatched to the sequential token-by-token nature of arithmetic coding. In the baseline transformer pipeline, 97\% of runtime is spent in the entropy coder. The target-interval capture optimisation (O3) removes this bottleneck, yielding an $18.96\times$ speedup and shifting runtime dominance entirely to the model forward pass.

These findings highlight that systems engineering matters as much as model quality. For a fixed model, throughput can vary by nearly two orders of magnitude depending on implementation. Techniques such as teacher-forced windows, pipelined multistream coding, and target-interval capture generalise beyond compression to other stateful sequential LLM workloads.

Despite these gains, LLM-based compression remains several orders of magnitude slower than LZ4 and zstd, making it impractical for latency-sensitive or high-throughput workloads. Its strongest use case today is cold, write-once data where storage reduction outweighs compute cost. The code dataset results $11.2\times$ compression versus zstd’s $3.06\times$ demonstrate the potential value for suitable data types.

Future work should focus primarily on reducing inference cost through smaller or specialised models, speculative decoding, and improved GPU scheduling. Since entropy coding is no longer the dominant bottleneck after optimisation, further gains are likely to come from faster inference rather than coding improvements.

Overall, LLM-based compression is not yet a replacement for classical codecs, but with careful systems engineering it is becoming practical for cold-storage analytical workloads where maximum compression ratio is more valuable than throughput.


\bibliographystyle{ACM-Reference-Format}
\bibliography{llm-compression}

@article{boncz20,
  author       = {Peter A. Boncz and
                  Thomas Neumann and
                  Viktor Leis},
  title        = {{FSST:} Fast Random Access String Compression},
  journal      = {Proc. {VLDB} Endow.},
  volume       = {13},
  number       = {11},
  pages        = {2649--2661},
  year         = {2020}
}

@inproceedings{adaptivecompression,
  author       = {Leon Windheuser and
                  Christoph Anneser and
                  Huanchen Zhang and
                  Thomas Neumann and
                  Alfons Kemper},
  title        = {Adaptive Compression for Databases},
  booktitle    = {{EDBT}},
  pages        = {143--149},
  publisher    = {OpenProceedings.org},
  year         = {2024}
}

@article{mosaic,
  author       = {Lukas Vogel and
                  Alexander van Renen and
                  Satoshi Imamura and
                  Viktor Leis and
                  Thomas Neumann and
                  Alfons Kemper},
  title        = {Mosaic: {A} Budget-Conscious Storage Engine for Relational Database
                  Systems},
  journal      = {Proc. {VLDB} Endow.},
  volume       = {13},
  number       = {11},
  pages        = {2662--2675},
  year         = {2020}
}

@inproceedings{transformer,
  author       = {Ashish Vaswani and
                  Noam Shazeer and
                  Niki Parmar and
                  Jakob Uszkoreit and
                  Llion Jones and
                  Aidan N. Gomez and
                  Lukasz Kaiser and
                  Illia Polosukhin},
  title        = {Attention is All you Need},
  booktitle    = {Advances in Neural Information Processing Systems 30: Annual Conference
                  on Neural Information Processing Systems 2017, December 4-9, 2017,
                  Long Beach, CA, {USA}},
  pages        = {5998--6008},
  year         = {2017},
  url          = {https://proceedings.neurips.cc/paper/2017/hash/3f5ee243547dee91fbd053c1c4a845aa-Abstract.html}
}

@article{lossless_data_compression_llm,
  title={Lossless data compression by large models},
  author={Li, Ziguang and Huang, Chao and Wang, Xuliang and Hu, Haibo and Wyeth, Cole and Bu, Dongbo and Yu, Quan and Gao, Wen and Liu, Xingwu and Li, Ming},
  journal={Nature Machine Intelligence},
  volume       = {7},
  number       = {5},
  pages        = {754--799},
  year         = {2025},
  url          = {www.nature.com/articles/s42256-025-01033-7},
  doi          = {https://doi.org/10.1038/s42256-025-01033-7}
}

@article{llmzip,
  author       = {Chandra Shekhara Kaushik Valmeekam and
                  Krishna Narayanan and
                  Dileep Kalathil and
                  Jean{-}Fran{\c{c}}ois Chamberland and
                  Srinivas Shakkottai},
  title        = {LLMZip: Lossless Text Compression using Large Language Models},
  journal      = {CoRR},
  volume       = {abs/2306.04050},
  year         = {2023},
  url          = {https://doi.org/10.48550/arXiv.2306.04050},
  doi          = {10.48550/ARXIV.2306.04050},
  eprinttype    = {arXiv},
  eprint       = {2306.04050}
}

@article{compression_llm_generated_data,
  author       = {Yu Mao and
                  Holger Pirk and
                  Chun Jason Xue},
  title        = {Lossless Compression of Large Language Model-Generated Text via Next-Token
                  Prediction},
  journal      = {CoRR},
  volume       = {abs/2505.06297},
  year         = {2025},
  url          = {https://doi.org/10.48550/arXiv.2505.06297},
  doi          = {10.48550/ARXIV.2505.06297},
  eprinttype    = {arXiv},
  eprint       = {2505.06297}
}

@article{finezip,
  author       = {Fazal Mittu and
                  Yihuan Bu and
                  Akshat Gupta and
                  Ashok Devireddy and
                  Alp Eren Ozdarendeli and
                  Anant Singh and
                  Gopala Anumanchipalli},
  title        = {FineZip : Pushing the Limits of Large Language Models for Practical
                  Lossless Text Compression},
  journal      = {CoRR},
  volume       = {abs/2409.17141},
  year         = {2024},
  url          = {https://doi.org/10.48550/arXiv.2409.17141},
  doi          = {10.48550/ARXIV.2409.17141},
  eprinttype    = {arXiv},
  eprint       = {2409.17141}   
}

@article{zstd,
  title={Zstandard Compression and the application/zstd Media Type},
  author={Yann Collet and Murray Kucherawy},
  journal={Internet Engineering Task Force (IETF) - RFC 8478},
  pages={1--54},
  year={2018},
  url={https://www.rfc-editor.org/rfc/pdfrfc/rfc8478.txt.pdf}
}

@article{fsst,
  author       = {Peter A. Boncz and
                  Thomas Neumann and
                  Viktor Leis},
  title        = {{FSST:} Fast Random Access String Compression},
  journal      = {Proc. {VLDB} Endow.},
  volume       = {13},
  number       = {11},
  pages        = {2649--2661},
  year         = {2020},
  url          = {http://www.vldb.org/pvldb/vol13/p2649-boncz.pdf}
}

@misc{pytorrent,
      title={{PyTorrent: A Python Library Corpus for Large-scale Language Models}}, 
      author={Mehdi Bahrami and N. C. Shrikanth and Shade Ruangwan and Lei Liu and Yuji Mizobuchi and Masahiro Fukuyori and Wei-Peng Chen and Kazuki Munakata and Tim Menzies},
      year={2021},
      eprint={2110.01710},
      archivePrefix={arXiv},
      primaryClass={cs.SE},
      url={https://arxiv.org/abs/2110.01710}, 
}

@inproceedings{goyalDeepZipLosslessData2019,
  title = {{{DeepZip}}: {{Lossless Data Compression Using Recurrent Neural Networks}}},
  shorttitle = {{{DeepZip}}},
  booktitle = {2019 {{Data Compression Conference}} ({{DCC}})},
  author = {Goyal, Mohit and Tatwawadi, Kedar and Chandak, Shubham and Ochoa, Idoia},
  date = {2019-03},
  pages = {575--575},
  issn = {2375-0359},
  doi = {10.1109/DCC.2019.00087},
  url = {https://ieeexplore.ieee.org/document/8712659},
}

@article{Huffman1952AMF,
  title = {A Method for the Construction of Minimum-Redundancy Codes},
  author = {Huffman, David A.},
  date = {1952},
  journaltitle = {Resonance},
  volume = {11},
  pages = {91--99},
  url = {https://api.semanticscholar.org/CorpusID:10606404}
}

@article{rissanenArithmeticCoding1979,
  title = {Arithmetic {{Coding}}},
  author = {Rissanen, J. and Langdon, G. G.},
  date = {1979-03},
  journaltitle = {IBM Journal of Research and Development},
  volume = {23},
  number = {2},
  pages = {149--162},
  issn = {0018-8646},
  doi = {10.1147/rd.232.0149},
  url = {https://ieeexplore.ieee.org/document/5390830},
}

@article{schmidhuberSequentialNeuralText1996,
  title = {Sequential Neural Text Compression},
  author = {Schmidhuber, J. and Heil, S.},
  date = {1996-01},
  journaltitle = {IEEE Transactions on Neural Networks},
  shortjournal = {IEEE Trans. Neural Netw.},
  volume = {7},
  number = {1},
  pages = {142--146},
  issn = {10459227},
  doi = {10.1109/72.478398},
  url = {http://ieeexplore.ieee.org/document/478398/},
}

@inproceedings{goyalDZipImprovedGeneralpurpose2021,
  title = {{{DZip}}: Improved General-Purpose Loss Less Compression Based on Novel Neural Network Modeling},
  shorttitle = {{{DZip}}},
  booktitle = {2021 {{Data Compression Conference}} ({{DCC}})},
  author = {Goyal, Mohit and Tatwawadi, Kedar and Chandak, Shubham and Ochoa, Idoia},
  date = {2021-03},
  pages = {153--162},
  issn = {2375-0359},
  doi = {10.1109/DCC50243.2021.00023},
  url = {https://ieeexplore.ieee.org/abstract/document/9418692},
}

@inproceedings{maoTRACEFastTransformerbased2022,
  title = {{{TRACE}}: {{A Fast Transformer-based General-Purpose Lossless Compressor}}},
  shorttitle = {{{TRACE}}},
  booktitle = {Proceedings of the {{ACM Web Conference}} 2022},
  author = {Mao, Yu and Cui, Yufei and Kuo, Tei-Wei and Xue, Chun Jason},
  date = {2022-04-25},
  series = {{{WWW}} '22},
  pages = {1829--1838},
  publisher = {Association for Computing Machinery},
  location = {New York, NY, USA},
  doi = {10.1145/3485447.3511987},
  url = {https://dl.acm.org/doi/10.1145/3485447.3511987},
  isbn = {978-1-4503-9096-5},
}

@inproceedings{liuDecMacDeepContext2019,
  title = {{{DecMac}}: {{A Deep Context Model}} for {{High Efficiency Arithmetic Coding}}},
  shorttitle = {{{DecMac}}},
  booktitle = {2019 {{International Conference}} on {{Artificial Intelligence}} in {{Information}} and {{Communication}} ({{ICAIIC}})},
  author = {Liu, Qian and Xu, Yiling and Li, Zhu},
  date = {2019-02},
  pages = {438--443},
  doi = {10.1109/ICAIIC.2019.8668843},
  url = {https://ieeexplore.ieee.org/document/8668843},
}

@misc{tacconelliNacrithNeuralLossless2026,
  title = {Nacrith: {{Neural Lossless Compression}} via {{Ensemble Context Modeling}} and {{High-Precision CDF Coding}}},
  shorttitle = {Nacrith},
  author = {Tacconelli, Roberto},
  date = {2026-02-24},
  eprint = {2602.19626},
  eprinttype = {arXiv},
  eprintclass = {cs.IT},
  doi = {10.48550/arXiv.2602.19626},
  url = {http://arxiv.org/abs/2602.19626},
}

@inproceedings{Kipf2026WaitingTD,
  title={Waiting to Decompress: The Economics of LLM-Based Compression},
  author={Andreas Kipf and Tobias Schmidt and Ping-Lin Kuo and Skander Krid and Moritz Rengert and Luca Heller and Andreas Zimmerer and Mihail Stoian and Varun Pandey and Alexander van Renen},
  booktitle={Conference on Innovative Data Systems Research},
  year={2026},
  url={https://api.semanticscholar.org/CorpusID:284161408}
}

@misc{touvronLLaMAOpenEfficient2023,
  title = {{{LLaMA}}: {{Open}} and {{Efficient Foundation Language Models}}},
  shorttitle = {{{LLaMA}}},
  author = {Touvron, Hugo and Lavril, Thibaut and Izacard, Gautier and Martinet, Xavier and Lachaux, Marie-Anne and Lacroix, Timothée and Rozière, Baptiste and Goyal, Naman and Hambro, Eric and Azhar, Faisal and Rodriguez, Aurelien and Joulin, Armand and Grave, Edouard and Lample, Guillaume},
  date = {2023-02-27},
  url = {https://arxiv.org/abs/2302.13971v1},
}

@inproceedings{kwonEfficientMemoryManagement2023,
  title = {Efficient {{Memory Management}} for {{Large Language Model Serving}} with {{PagedAttention}}},
  booktitle = {Proceedings of the 29th {{Symposium}} on {{Operating Systems Principles}}},
  author = {Kwon, Woosuk and Li, Zhuohan and Zhuang, Siyuan and Sheng, Ying and Zheng, Lianmin and Yu, Cody Hao and Gonzalez, Joseph and Zhang, Hao and Stoica, Ion},
  year = 2023,
  month = oct,
  series = {{{SOSP}} '23},
  pages = {611--626},
  publisher = {Association for Computing Machinery},
  address = {New York, NY, USA},
  doi = {10.1145/3600006.3613165},
  urldate = {2026-05-21},
  isbn = {979-8-4007-0229-7},
}

@inproceedings{zhengSGLangEfficientExecution2024,
  title = {{{SGLang}}: Efficient Execution of Structured Language Model Programs},
  shorttitle = {{{SGLang}}},
  booktitle = {Proceedings of the 38th {{International Conference}} on {{Neural Information Processing Systems}}},
  author = {Zheng, Lianmin and Yin, Liangsheng and Xie, Zhiqiang and Sun, Chuyue and Huang, Jeff and Yu, Cody Hao and Cao, Shiyi and Kozyrakis, Christos and Stoica, Ion and Gonzalez, Joseph E. and Barrett, Clark and Sheng, Ying},
  year = 2024,
  month = dec,
  series = {{{NeurIPS}} '24},
  volume = {37},
  pages = {62557--62583},
  isbn = {979-8-3313-1438-5}
}

@article{szlangWorkloadInsightsSnowflake2025,
  title = {Workload {{Insights}} from the {{Snowflake Data Cloud}}: {{What Do Production Analytic Queries Really Look Like}}?},
  shorttitle = {Workload {{Insights}} from the {{Snowflake Data Cloud}}},
  author = {Szlang, Jan Vincent and Bress, Sebastian and Cattes, Sebastian and Dees, Jonathan and Funke, Florian and Heimel, Max and Oleynik, Michel and Oukid, Ismail and Maltenberger, Tobias},
  date = {2025-08},
  journaltitle = {Proceedings of the VLDB Endowment},
  shortjournal = {Proc. VLDB Endow.},
  volume = {18},
  number = {12},
  pages = {5126--5138},
  issn = {2150-8097},
  doi = {10.14778/3750601.3750632},
  url = {https://dl.acm.org/doi/10.14778/3750601.3750632},
}

@article{vanrenenWhyTPCNot2024,
  title = {Why {{TPC}} Is {{Not Enough}}: {{An Analysis}} of the {{Amazon Redshift Fleet}}},
  shorttitle = {Why {{TPC}} Is {{Not Enough}}},
  author = {Van Renen, Alexander and Horn, Dominik and Pfeil, Pascal and Vaidya, Kapil and Dong, Wenjian and Narayanaswamy, Murali and Liu, Zhengchun and Saxena, Gaurav and Kipf, Andreas and Kraska, Tim},
  date = {2024-07},
  journaltitle = {Proceedings of the VLDB Endowment},
  shortjournal = {Proc. VLDB Endow.},
  volume = {17},
  number = {11},
  pages = {3694--3706},
  issn = {2150-8097},
  doi = {10.14778/3681954.3682031},
  url = {https://dl.acm.org/doi/10.14778/3681954.3682031},
}

@inproceedings{Lam2015NumbaAL,
  title={Numba: a LLVM-based Python JIT compiler},
  author={Siu Kwan Lam and Antoine Pitrou and Stanley Seibert},
  booktitle={Proceedings of the Second Workshop on the LLVM Compiler Infrastructure in HPC},
  pages={1--6},
  year={2015},
  doi={10.1145/2833157.2833162},
  url={https://dl.acm.org/doi/10.1145/2833157.2833162}
}

\clearpage

\appendix
\section{Engine Backend Implementation Details}
\label{app:engine-backends}

\subsection{vLLM Logits Extraction}

Since vLLM does not return next-token scores in a row-stable order suitable
for compression, early versions used \texttt{prompt\_logprobs}'s output
for rank-based encodings and experimented with bulk rank extraction from that
interface. This is sufficient for bitpacked and Huffman variants, where only the
rank of the true token in the sorted score list is needed, but arithmetic coding
requires a full distribution over the reduced vocabulary at every step, making
top-$k$ logprob APIs insufficient.

We also evaluated vLLM's \texttt{logits\_processors} interface with score
capture to temporary files. This confirmed that vLLM can expose the required
scores, but the file handoff introduced latency that made the path uneconomical
for the compression workload.

The current implementation uses POSIX shared memory. The host process allocates
a capture buffer and passes its name through environment variables. The logits
processor attaches to that buffer, copies the selected logits as
\texttt{float32}, and marks completed row-step entries with readiness flags.
Each request carries a row identifier in vLLM's sampling parameters so the host
can recover the original batch order even when vLLM reorders requests
internally.

\subsection{TensorRT-LLM Tensor Extraction}

TensorRT-LLM can return logits in several payload formats and with
engine-specific padding in the vocabulary dimension. The predictor extracts the
last valid time step from the returned tensors, trims any padded vocabulary
columns back to the semantic tokenizer vocabulary, and then applies the global
token mask. We prefer lower-level model-runner interfaces when available and
fall back to \texttt{LLM.generate} otherwise; both are normalised to the same
$[\mathrm{batch}, \mathrm{vocab}]$ contract.

\section{Optimization Implementation Details}
\label{app:optimizations}

\subsection{Teacher-Forced Windows}

The logits processor forces the sampler to emit the known target token at each
internal decode step by setting all other logits to $-\infty$. The window size
$w$ is bounded by the configured context length and the number of retained
context tokens so that every captured score vector corresponds exactly to the
prefix the decoder will reconstruct.

In our experience this optimization is not compatible with vLLM's tensor
parallelism and is therefore restricted to single-GPU configurations. Even at
large batch sizes, the overhead of POSIX shared-memory synchronisation remains
higher than the transformer path. The optimization nonetheless illustrates how
the compression workload can benefit from custom scheduling and data-movement
strategies inside the inference engine.

\subsection{Multistream Arithmetic Coding}

Probabilities are normalized in \texttt{float64}, clamped to non-negative
values, and scaled to a fixed total $T = 2^{18}$ with a minimum count of one
per symbol. The target interval $(\ell, h, T)$ is extracted from row-wise
prefix sums without constructing a per-row cumulative table on the host.

The Numba-compiled~\cite{Lam2015NumbaAL} backend implements the standard integer
range-coder update loop with renormalization and underflow handling, operating
on compact arrays of interval triples rather than Python objects. A packed
variant writes bits directly into byte arrays and returns an explicit bit count,
avoiding the memory overhead of materializing Python lists. Row streams are
independent, so final encoding can be parallelised across CPU threads.

On the transformer backend, interval conversion and batch encoding for step $t$
are submitted to a background thread immediately after the probability batch is
produced. The main thread begins step $t{+}1$ without waiting and synchronises
only if the background task for step $t$ has not yet completed, overlapping GPU
execution with CPU-side coding latency.

\subsection{Target-Interval Capture}

In target-interval mode the inference backend reduces the vocabulary if
requested, computes the softmax locally, derives the interval for the forced
token, and writes only a fixed-size record $(\ell, h, T, p_{\mathrm{target}})$
into shared memory. This changes the data-transfer cost from $O(|V'|)$ floats
per row-step to $O(1)$, greatly reducing PCIe and shared-memory traffic and
removing the host-side cumulative-table construction step for the transformer
and TensorRT-LLM paths.

\end{document}